\documentclass[prb,aps,twocolumn,showpacs,preprintnumbers,amsmath,amssymb,superscriptaddress]{revtex4-1}
\usepackage{graphicx}
\usepackage{color}
\usepackage{bm}% bold math
\usepackage{epstopdf}
\usepackage{color}

\begin{document}

\title{Negative thermal expansion and antiferromagnetism in the actinide oxypnictide NpFeAsO}

\author{T. Klimczuk}
\email[]{Tomasz.Klimczuk@ec.europa.eu, }
\affiliation{European Commission, JRC, Institute for Transuranium Elements, Postfach 2340, 76125 Karlsruhe, Germany}
\affiliation{Faculty of Applied Physics and Mathematics, Gdansk University of Technology, Narutowicza 11/12, 80-952 Gdansk, Poland}

\author{H. C. Walker}
\affiliation{European Synchrotron Radiation Facility, 6 rue Jules Horowitz, BP220, 38043 Grenoble Cedex 9, France}
\affiliation{Deutsches Elektronen-Synchrotron (Hasylab at DESY), 22607 Hamburg, Germany}

\author{R. Springell}
\affiliation{London Centre for Nanotechnology and Department of Physics and Astronomy, University College London, London WC1E 6BT, United Kingdom}
\affiliation{Royal Commission for the Exhibition of 1851 Research Fellow, Interface Analysis Centre, University of Bristol, Bristol BS2 8BS, United Kingdom}

\author{A. B. Shick}
\affiliation{European Commission, JRC, Institute for Transuranium Elements, Postfach 2340, 76125 Karlsruhe, Germany}

\affiliation{Institute of Physics, ASCR, Na Slovance 2, 18221 Prague 8, Czech Republic}

\author{A. H. Hill}
\altaffiliation[Currently at: ]{Johnson Matthey Technology Centre, Sonning Common, UK}
\affiliation{European Synchrotron Radiation Facility, 6 rue Jules Horowitz, BP220, 38043 Grenoble Cedex 9, France}

\author{P. Gaczy\'nski}
\affiliation{European Commission, JRC, Institute for Transuranium Elements, Postfach 2340, 76125 Karlsruhe, Germany}

\author{K. Gofryk}
\affiliation{Los Alamos National Laboratory, Los Alamos NM, USA}

\author{S. A. J. Kimber}
\affiliation{European Synchrotron Radiation Facility, 6 rue Jules Horowitz, BP220, 38043 Grenoble Cedex 9, France}

\author{C. Ritter}
\affiliation{Institute Laue-Langevin, 6 rue Jules Horowitz, BP156, 38042 Grenoble Cedex 9, France}

\author{E. Colineau}
\author{J.-C. Griveau}
\author{D. Bou\"{e}xi\`{e}re}
\author{R. Eloirdi}
\affiliation{European Commission, JRC, Institute for Transuranium Elements, Postfach 2340, 76125 Karlsruhe, Germany}

\author{R. J. Cava}
\affiliation{Department of Chemistry, Princeton University, Princeton NJ 08544, USA }

\author{R. Caciuffo}
\affiliation{European Commission, JRC, Institute for Transuranium Elements, Postfach 2340, 76125 Karlsruhe, Germany}

\date{\today}

\begin{abstract}
A neptunium analogue of the LaFeAsO tetragonal layered compound has been synthesized and characterized by a variety of experimental techniques. The occurrence of long-range magnetic order below a critical temperature $T_\mathrm{N}$ = 57 K is suggested by anomalies in the temperature-dependent magnetic susceptibility, electrical resistivity, Hall coefficient, and specific heat curves. Below $T_\mathrm{N}$, powder neutron diffraction measurements reveal an antiferromagnetic structure of the Np sublattice, with an ordered magnetic moment of 1.70 $\pm$ 0.07 $\mu_B$ aligned along the crystallographic $c$-axis. No magnetic order has been observed on the Fe sublattice, setting an upper limit of about 0.3 $\mu_B$ for the ordered magnetic moment on the iron. High resolution x-ray powder diffraction measurements exclude the occurrence of lattice transformations down to 5 K, in sharp contrast to the observation of a tetragonal-to-orthorhombic distortion  in the rare-earth analogues, which has been associated with the stabilization of a spin density wave on the iron sublattice. Instead, a significant expansion of the NpFeAsO lattice parameters is observed with decreasing temperature below $T_\mathrm{N}$, corresponding to a relative volume change of about 0.2$\%$ and to an invar behavior between 5 and 20 K. First-principle electronic structure calculations based on the local-spin density plus Coulomb interaction and the local density plus Hubbard-I approximations provide results in good agreement with the experimental findings.
\end{abstract}
\maketitle

\section{Introduction}

Rarely in physics has a single family of compounds generated as much interest as the iron-pnictide superconductors \cite{Zaanen,Grant,Kivelson,Norman,Pickett}. Research in this field is being conducted at such a hectic pace that no fewer than 2500 articles have cited the paper that reported the discovery of superconductivity in $\mathrm{LaFeAsO}_{1-x}\mathrm{F}_{x}$ \cite{Kamihara2008}. The reason for this frenzy of activity is clear: the superconducting transition temperatures are very high, second only to the cuprates.

%What is the upper limit for $\mathrm{\textit{T}_{C}}$? Are the magnetic interactions strongly correlated and local, or weakly correlated and itinerant? How are they connected to the observation of superconductivity?

The replacement of the rare-earth (\textit{R}) species in the \textit{R}FeAsO ``1111'' phase with Np, without change to the room temperature structure, and the ability to subsequently replace Np with heavier actinides: Pu and Am, represents a real possibility to control correlations (from weak to strong, Np to Am) and to potentially extend the already rich physics of the iron pnictides.

The parent \textit{R}FeAsO ``1111'' compounds have so far involved rare-earth $\mathrm{3^{+}}$ ions in the oxide layers. These have ionic radii $\mathrm{\sim1\,\AA}$ and have partially filled \textit{4f} shells, which are atomic-like in nature. The influence of the oxide layer, in the behavior of these materials, beyond being a `spacer', is a seldom addressed question. In fact, from a materials point of view this is a difficult problem to investigate, since elements that possess similar chemical properties to the rare earths, both in terms of bonding and size, are hard to find. For ideal candidates one needs to descend to the bottom of the periodic table, specifically, the transuranic actinide elements, Np, Pu and Am. These have ionic radii of almost identical magnitude to the majority of the lanthanides, and are commonly present in the trivalent state in similar compounds. Not only do they represent an alternative to the lanthanides \textit{per se}, but they can also be used to study the effect of the degree of localization of the \textit{f} electron states (in this case \textit{5f}) on the observed properties. This is a commonly adopted strategy in the actinide community when investigating new compounds; the behavior of the actinide metals as one traverses the period from uranium to americium changes from itinerant, transition metal-like electronic behavior to localized, atomic-like behavior.

Here, we report on the synthesis and characterization of NpFeAsO, an isostructural Np analogue of LaFeAsO. Magnetic, electrical transport, and specific heat measurements have been used to characterize its macroscopic physical properties, whereas neutron and x-ray diffraction have been used to monitor the evolution of the magnetic and crystallographic structures with decreasing temperature. The experiments show the development of antiferromagnetic order on the Np sublattice below $T_\mathrm{N}$ = 57 K, and the absence of observable crystallographic distortions down to 5 K. This is at variance with the behaviour of the rare-earth analogues \textit{R}FeAsO, where the occurrence of a tetragonal-to-orthorhombic lattice transformation is observed well above the critical temperature of the rare-earth sublattice. On the other hand, the magnetic transition in NpFeAsO is accompanied by a large expansion of the unit cell volume below $T_\mathrm{N}$ followed by an invar behavior below about 20 K. First-principle calculations of the electronic structure provide a Fermi surface with almost 2-dimensional character, very similar to the one previously reported for \textit{R}FeAsO, and predict an antiferromagnetic ground state with a staggered magnetization along the tetragonal $c$-axis. The anomalous thermal expansion is qualitatively described within a local-density plus Hubbard-I approximation.

\section{Experimental}

\begin{figure*}[ht]
\centering
\includegraphics[width=0.9\textwidth,bb=10 120 580 415,clip]{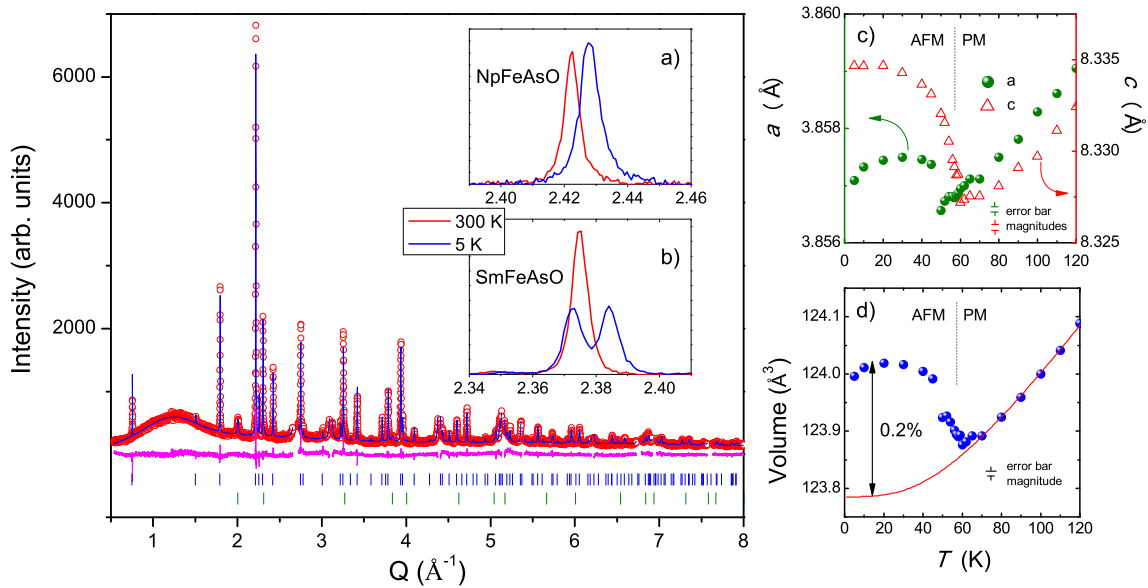}
\caption{(Color online) Fit (solid blue line) of the refined structural model (Rietveld method) to the high-resolution room-temperature x-ray powder diffraction data (red circles) for NpFeAsO, confirming the tetragonal $P4/nmm$ ZrCuSiAs-type structure. The data were collected on the ID31 (ESRF) high resolution x-ray powder diffractometer with a wave-length of $\lambda=0.35$~\AA. Tick marks are shown for two refined phases: NpFeAsO (upper, blue) and an $8\%$ NpO$_2$ impurity (lower, green). Insets (a) and (b) make a comparison of data for NpFeAsO and SmFeAsO \cite{Maroni}, where only SmFeAsO shows the splitting of the room temperature (red) tetragonal ($P4/nmm$) (111) Bragg reflection into the orthorhombic ($Cmma$) (201) and (021) reflections at 5 K (blue). Panels (c) and (d) show the variation in the $a$ and $c$ lattice parameters, and of the unit cell volume as a function of temperature, as extracted from the x-ray powder diffraction measurements, revealing a marked negative thermal expansion below $T\simeq60$~K in NpFeAsO. The red line through the volume data is a fit to the second order of the Gr\"{u}neisen approximation for the zero pressure equation of state, as discussed in the text.\label{fig:synch}}
\end{figure*}

The first and not inconsiderable hurdle one has to overcome when studying these materials is sample synthesis. This is non-trivial in the case of the radioactive and toxic actinides. One point of caution at this stage is that we have started our study with Np, whereas it is most common to begin with U. For the case of the oxypnictides, replacing the rare-earth element by U is highly unlikely to be successful, given that U can be either tri- or tetravalent, and is the latter in the vast majority of oxygen containing U-based compounds.

Polycrystalline samples of $\mathrm{NpFeAsO}$ were synthesized at the Institute for Transuranium Elements (ITU) by solid state reaction using $\mathrm{Fe_{3}O_{4}}$ (Alfa Aesar 99.997\%), elemental Fe (Alfa Aesar 99.998\%), and crystals of NpAs as starting materials. The thoroughly mixed powder was pressed into a pellet, sealed in an evacuated silica ampoule and heated at $\mathrm{900\,^{\circ}C}$ for 48 hours, before finally being furnace quenched. All operations were carried out in a radioprotected glovebox with low oxygen and water concentrations.

High resolution X-ray diffraction data were measured as a function of temperature ($5\leq T\leq300$~K) on the powder diffraction beamline, ID31 at the European Synchrotron Radiation Facility (ESRF), France, using an incident beam energy of $35$~keV. Powderized sample for this experiment was doubly encapsulated, being put inside a kapton tube (50 $\mu$m) and inserted in a plexiglass hollow cylinder with a wall thickness of 200 $\mu$m.

The magnetic susceptibility measurements were performed at ITU on a Quantum Design MPMS-7 SQUID magnetometer, in applied magnetic fields $\mu_0 H_1$ = 5 T and $\mu_0 H_2$ = 7 T, where we define $\chi_{DC}=[M(H_2)-M(H_1)]/(H_2-H_1)$. Electrical resistivity, Hall effect and heat capacity were measured using a Quantum Design Physical Properties Measurement System (PPMS-9). The electrical resistivity was determined using a standard 4-probe DC technique, with four 0.05 mm diameter platinum wires glued to the sample using silver epoxy (Epotek H20E). The Hall resistance ($R_{H}$) was determined by voltage measurements $V_{H}$ under applied magnetic field at +9 and -9T. The field response $V_{H}$(B) at fixed temperatures has been measured to confirm results obtained when ramping in temperature. The heat capacity was measured using a standard relaxation calorimetry method, where the data were corrected for the contribution of the Stycast encapsulation material (2850 FT) by using an empirical relation determined previously.

Neutron powder diffraction was performed on the high-intensity two-axis D20 diffractometer at the Institut Laue Langevin (ILL), France, in the high flux mode, with an incident wave-length of $2.42$~\AA, using a (002) pyrolitic graphite monochromator. About 250 mg of NpFeAsO powder was used, within a triple-walled Al container. Data were collected at 5, 30, 90, 190, and 290 K using the ILL trans-uranic samples orange cryostat. Variable temperature ramp data were also collected between 30 and 90 K.

%The $\mathrm{^{237}Np}$ M\"{o}ssbauer spectra were measured using a transmission geometry spectrometer with a $\mathrm{100\,mCi}$ $\mathrm{^{241}Am}$ source at $T\simeq4$~K, on a fine powder sample with an optimal thickness of $\mathrm{140\,mg\,Np\,cm^{-2}}$. The isomer shift (IS) is given relative to $\mathrm{NpAl_{2}}$ at $\mathrm{\sim4\,K}$). The data are analysed using Lorentzian lineshapes. The position and relative intensities of the absorption lines were calculated by solving the complete Hamiltonian for the hyperfine interactions in both the excited and ground nuclear state of $\mathrm{^{237}Np}$.

%
\section{Results and discussion}

\begin{figure*}[t]
\centering
\includegraphics[width=0.8\textwidth,bb=5 5 705 545,clip]{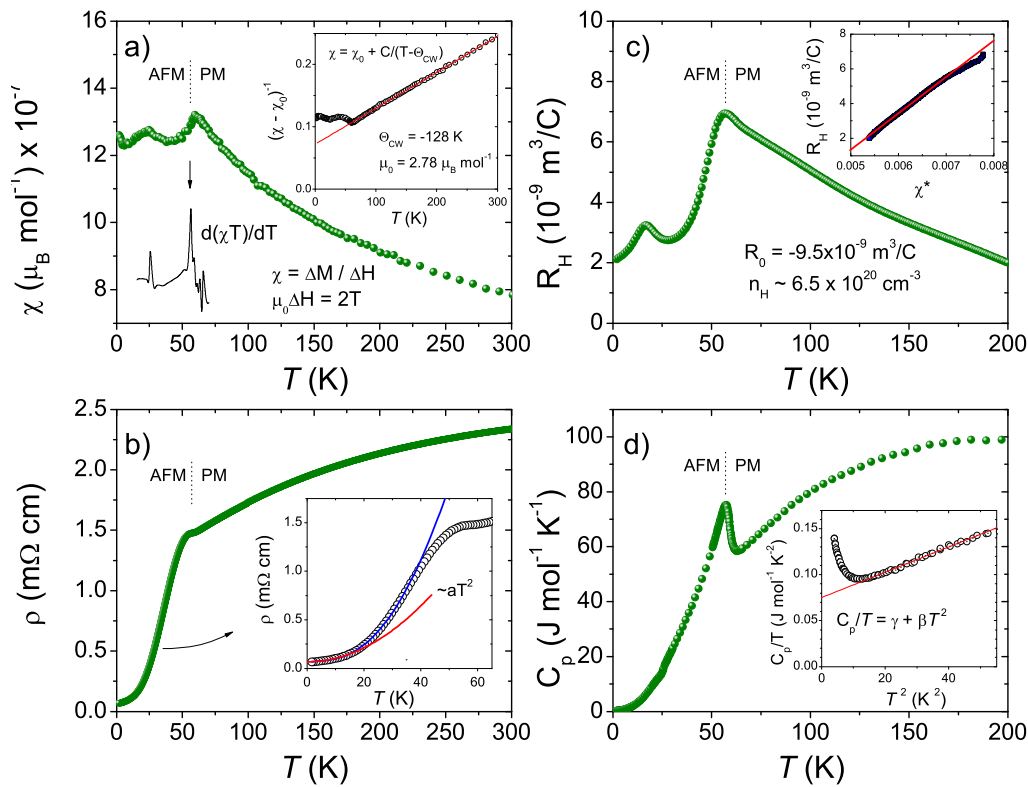}
\caption{(Color online)Bulk physical properties of NpFeAsO. (a) The magnetic susceptibility $\chi=\Delta M/\Delta H$ data (closed circles) and $d(\chi\,T)/dT$ (solid line). The inset shows a Curie-Weiss fit to the inverse susceptibility from which an effective paramagnetic moment of $2.78\pm0.06\mu_\mathrm{B}$ and a Curie-Weiss temperature of $-128\pm2$~K are obtained. (b) The electrical resistivity measured in $H=0$~T. The inset highlights the magnetically ordered region, showing two fits to the data, indicating that the low temperature resistivity does not vary simply as $T^2$ (red line), but instead an improved fit is obtained on including an antiferromagnetic gap (blue line). (c) The Hall coefficient as a function of temperature. The inset shows the least squares fit described in the text. (d) The heat capacity, including an inset showing a fit to the low temperature data.}
\label{fig:props}
\end{figure*}

Upon the successful synthesis of NpFeAsO, our initial step was to check the crystal structure and sample purity using x-ray powder diffraction. At room temperature this indicates that this Np compound is an isostructural analogue to the celebrated rare-earth iron oxypnictide family \textit{R}FeAsO, as demonstrated by the successful refinement of the high resolution data shown in Figure~\ref{fig:synch} to the ZrCuSiAs-type structure. Extra lines appearing in the diffraction pattern correspond to NpO$_2$, which is the only observable impurity phase (about $8\%$ in weight). On cooling to $5$~K no orthorhombic distortion was observed within the resolution of our data (inset (a)), in contrast to all of the \textit{R}FeAsO compounds (such as SmFeAsO in inset (b)). The absence of the crystallographic distortion immediately prompts a question regarding the nature of the magnetism in NpFeAsO, since it is believed that in the rare-earth compounds the distortion is intrinsically related to the development of a spin density wave on the iron, breaking the rotational symmetry within the plane. Inspection of Figure~\ref{fig:props} reveals that bulk thermodynamic magnetic and transport property measurements give no indication of an anomaly within the standard temperature range for the \textit{R}FeAsO spin density wave ordering ($120-140$~K). Instead, anomalies are only observed at 25 K, corresponding to the phase transition in the NpO$_2$ impurity phase \cite{caciuffo87}, and significantly at $\simeq60$~K.%, the temperature corresponding to the onset of the anomalous thermal expansion.

Even more striking and unusual, however, is the observed variation of the lattice parameters as a function of temperature. As shown in Figures~\ref{fig:synch}(c) and (d), the standard thermal expansion behavior stops abruptly below $T\sim60$~K, and as the temperature is decreased the crystallographic cell expands. The very accurate data obtained at the ESRF ID31 high-resolution powder diffraction beamline allows us to study the effect along both principal crystallographic directions in detail. Although the temperature dependences for both $a$ and $c$ lattice parameters, show clear minima, $T_{min}$, the $c$-axis minimum is about 10 K higher than in the $a$-axis direction. Below $T_{min}$ the $c$ lattice parameter increases monotonically and saturates for $T < 20K$. On the contrary, the $a$ lattice parameter reaches a maximum at around 30 K and then starts to decrease with decreasing temperature. The relative lattice size change differs by a factor of 4, the one in the $c$-axis direction being the greater. The above conclusions are not affected by the presence of the NpO$_2$ impurity, whose lattice parameter shows a regular temperature behavior with a contraction at the onset of electric quadrupole order (T$_0$ = 25 K) \cite{Mannix99}.

Using the second order Gr\"{u}neisen approximation for the zero pressure equation of state\cite{Wallace,Vocadlo}, the volumetric data was modeled according to:

\begin{equation}
V(T)=\frac{V_0 U}{Q-bU}+V_0,
\end{equation}
where $U$ is the internal energy, which we have calculated following the Debye approximation:

\begin{equation}
U(T)=9Nk_BT\left(\frac{T}{\Theta_D}\right)^3\int_0^{\Theta_D/T}\frac{x^3dx}{e^x-1},
\end{equation}
$N$ is the number of atoms in the unit cell, $k_B$ is the Boltzmann constant, and $\Theta_D$ is the Debye temperature, obtaining
$\Theta_D=287$~K, $Q=0.52$~eV, $b=260$ and $V_0=123.78$~\AA$^3$, and hence a negative thermal expansion of $\omega = \Delta V / V=0.2\%$.

The presence and form of the anomaly in the magnetic susceptibility (Fig~\ref{fig:props}(a)) at $\sim57$~K, implies the onset of antiferromagnetic ordering, and therefore that the anomalous thermal expansion is a consequence of a strong coupling between the magnetism and the lattice. In order to estimate the precise value of the N\'{e}el temperature, we followed Ref. \cite{Dutta}, and plotted $d(\chi\,T)/dT$ (solid line) on the same temperature scale. The maximum of $d(\chi\,T)/dT$ is observed at $T_\mathrm{N}=57$~K. Above $\sim$ 100K, the susceptibility curve has a Curie-Weiss behavior, with an effective paramagnetic moment $\mu_\mathrm{eff}=2.78\pm0.06~\mu_\mathrm{B}$ and a Curie-Weiss temperature $\Theta_\mathrm{CW}=-128\pm2$~K. The above values have been obtained by fitting the inverse susceptibility curve to a straight line, taking into account the contribution of the NpO$_2$ impurity phase by scaling the data reported in Ref.~\onlinecite{caciuffo03a}.

In the intermediate coupling scheme, $\mu_\mathrm{eff}$ of Np$^{3+}$ is $2.755~\mu_\mathrm{B}$, and hence our results imply that the neptunium is trivalent, as is the case for the rare-earth species in the ReFeAsO isostructural compounds. The magnetic moment on the iron is likely to be small given that the majority of the magnetism can be attributed to the Np moments. The negative Curie-Weiss temperature suggests the presence of antiferromagnetic interactions. An additional anomaly is present in the data at $T=25$~K. This corresponds to the magnetic triakontadipole phase transition of NpO$_2$ \cite{santini06,magnani08,santini09}, which is present as an impurity phase in our sample, as detected in the x-ray diffraction pattern.

Figure \ref{fig:props}(b) shows the resistivity, revealing a sharp drop at the magnetic transition. The low temperature resistivity can be described by the presence of AF interactions at high temperatures with the opening of a magnetic gap near the magnetic transition temperature \cite{felectrons}, as demonstrated by the fit in the insert to the function (blue solid line):
\begin{equation}
\rho(T)=\rho_{0}+aT^2+bT(1+2T/\Delta)\exp^{-\Delta/T},
\end{equation}
where $\mathrm{\rho_{0}=70\pm1\,\mu\Omega cm}$, $\mathrm{\textit{a}=0.27\pm0.01\,\mu\Omega cm K^{-2}}$, $\mathrm{\textit{b}=20\pm1\,\mu\Omega cm K^{-1}}$ and $\mathrm{\Delta=49\pm1}$~K. The value of $\mathrm{\Delta}$ is comparable to that obtained for $\mathrm{NpCoGa_{5}}$ ($\mathrm{\Delta=55\,K}$), which orders antiferromagnetically at $T_\mathrm{N}=47$~K \cite{Colineau}.

Whilst the electrical resistivity indicates that NpFeAsO is a reasonable metal, the number and nature of the carriers were further investigated by means of the Hall Effect. The temperature dependence of the Hall effect of NpFeAsO is shown in Fig.~\ref{fig:props}(c). At room temperature the Hall coefficient $R_{H}$ is positive and of the order of 2$\times$10$^{-9}$ m$^{3} C^{-1}$. It increases with decreasing temperature down to the N\'{e}el temperature $T_\mathrm{N}=57$~K where $R_{H}(T)$ exhibits a distinct maximum, reminiscent of the maximum in the temperature dependence of the magnetic susceptibility (see Fig.~\ref{fig:props}(a)). It has been shown experimentally and theoretically, based on the Anderson periodic model with the crystal electric field effect, that in magnetic materials the Hall coefficient may be described as a sum of two parts:
\begin{equation}
R_{H}(T) = R_{0}+R_{s}\chi^{*}(T),
\end{equation}
where $R_{0}$ is the normal Hall effect due to the Lorenz motion of carriers and the second term, the anomalous Hall effect, is related to the magnetic scattering\cite{Putley,Kontani,Nagaosa}. The reduced susceptibility $\chi^{*}(T)$ is approximated by $\chi(T)/C$ where $C$ is the Curie-Weiss constant\cite{Kontani}. In the case of NpFeAsO, a least squares fitting of the above equation to the experimental data in the temperature range $60-200$~K (see the inset to Fig.~\ref{fig:props}(c)) resulted in the values $R_{0}=-9.5\times10^{-9}$ m$^{3} C^{-1}$ and $R_{s}=2.1\times10^{-6}$ m$^{3} C^{-1}$ for the normal and anomalous Hall coefficients. The anomalous Hall effect, likely caused by the magnetism of the Np ions, is dominant, and therefore a positive $R_{H}$ is observed. However, the negative value of $R_{0}$ indicates that electrons are the dominant carriers, most probably due to their higher mobility. The single band model provides an estimate for the concentration of free electrons to be $6.5\times10^{20}~$cm$^{-3}$, which should be considered as the upper limit of the actual carrier concentration in NpFeAsO. Interestingly, the values of the ordinary Hall coefficient, as well as its sign, and carrier concentration obtained for NpFeAsO are similar to the ones derived for similar \textit{R}FeAsO systems based on the lanthanide elements\cite{Sefat,McGuire,Suzuki,Liu}. Neither resistivity, nor Hall effect measurements on NpFeAsO show anomalies as observed around 150 K in transport properties measurements of \textit{R}FeAsO (\textit{R} = La, Ce, Pr, Nd)\cite{McGuire}.

Figure~\ref{fig:props}(d) presents the heat capacity data measured down to 2K. Two anomalies are clearly identified, namely a sharp $\lambda$-type cusp at $\sim$60 K, that coincides with the onset of antiferromagnetic order in NpFeAsO, and a shoulder at 25K, associated with the multipolar phase transition of NpO$_2$ impurity \cite{magnani05}. The insert shows a fit to $C_P/T=\gamma+\beta T^2$ in the $3.5 - 7$~K temperature range, where $\gamma T$ is the electronic contribution to the heat capacity, and $\beta T^3$ ($\beta=12\pi^4Nk_B/5\theta_D^3$) is the acoustic phonon contribution in the low temperature limit of the Debye model, where $\theta_D$ is the Debye temperature, $N$ is the number of atoms per formula unit, and $k_B$ is the Boltzmann constant. The contribution of the NpO$_2$ impurity to the specific heat data have been subtracted by scaling the data reported for NpO$_2$ in ref. \onlinecite{magnani05}.

In the magnetically ordered state the magnetic specific heat term ($C_{mag}$) should be taken into account. In ref. \onlinecite{Colineau}, the formula that describes the specific heat of magnons with an energy gap $\Delta$ in their dispersion relation, $C_{mag}=\alpha T^{1/2}\exp{(-\Delta/T)}$, was succesfully used in order to fit $C_{mag}$ of NpCoGa$_5$. Using the value of $\Delta$ = 49 K obtained from the resistivity fit, and taking $\alpha$ = 3.7 ~J mol$^{-1}$K$^{-3/2}$, as reported for  NpCoGa$_5$, $C_{mag}$ term can be estimated. In the 3.5 K to 7 K temperature range, $C_{mag}$ changes from $\mathrm{6\,\mu J mol^{-1}}$K$^{-1}$ to $\mathrm{9\,mJ mol^{-1}}$ K$^{-1}$, which is 5 and 2 orders of magnitude smaller than the measured value of $C_{P}$ and does not influence the fit shown in the inset of Figure~\ref{fig:props}(d), where no deviations from linear behavior are observed. The values of $\gamma$ and $\Theta_D$ extracted from the fit are $75\pm1$~mJ mol$^{-1}$K$^{-2}$, and $178\pm2$~K respectively. The obtained Debye temperature is not in good agreement with the value obtained from the fit of volumetric data. The observed upturn in $C_P/T$ below $T=3$~K is caused by the nuclear Schottky anomaly, which is commonly observed in Np compounds\cite{Klimczuk,Colineau}. In order to calculate the magnetic entropy, the phonon contribution to the specific heat must be subtracted. However, we can not use reported data for parent \textit{R}FeAsO because all of them undergo a structural phase transition such that at low temperature they posses the orthorhombic crystal structure. According to our knowledge, despite of a lot of effort, synthesis of UFeAsO and ThFeAsO has not been successful. Therefore, the entropy removed upon magnetic ordering cannot be calculated at this moment.

Given both $\mathrm{\gamma}$ and $\mathrm{\chi(0~K)}$, which was obtained from the interpolation of the $\mathrm{C-W}$ fit to $T=0$~K, the Wilson ratio,
\begin{equation}
\mathrm{R_{W}=(\pi k_{B})^{2}\chi(0 K)/3\mu_{eff}^{2}\gamma},
\end{equation}
can be estimated for NpFeAsO, yielding a value for $\mathrm{R_{W}}$ of 1.3. This is slightly higher than the free electron value ($\mathrm{R_{W}=1}$) and less than expected for heavy fermion compounds ($\mathrm{R_{W}=2}$).

%In order to identify the nature of the magnetic ordering, evidenced by the anomalies in the bulk properties at $T\sim57$~K, $^{237}$Np M\"{o}ssbauer spectroscopy and powder neutron diffraction experiments were undertaken. At $T=60$~K the M\"{o}ssbauer spectrum (Fig.~\ref{fig:mag_order}(a)) shows one strong narrow absorption line, indicating that at this temperature NpFeAsO is in the paramagnetic state, and one weaker absorption peak, arising due to the NpO$_2$ impurity phase. The spectrum demonstrates that the Np ions only occupy a single crystallographic site, and the narrow line-width excludes the possibility of any chemical disorder at this site. A fit to the spectrum gives an isomer shift of $9.09\pm0.01$~mm s$^{-1}$ with respect to NpAl$_2$, consistent with the range expected for an Np$^{3+}$ ion in conducting materials, thus confirming the valence result from the Curie-Weiss fit to the inverse susceptibility. Below $57$~K, a magnetic splitting of the spectrum is observed. The data can be analysed using a unique set of hyperfine parameters with a magnetic hyperfine field $B_{hf}=372\pm2$~T collinear with the main component of the electric field gradient. This fact indicates that the Np magnetic moments are oriented along the c-axis. Using the relationship $B_{hf}/\mu_\mathrm{Np}=215\pm5\, \mathrm{T}/\mu_\mathrm{B}$, this corresponds to a saturated ordered Np magnetic moment of $1.73\pm0.05\mu_\mathrm{B}$, oriented along the $c$-axis.

\begin{figure}[t]
\centering
\includegraphics[width=0.45\textwidth]{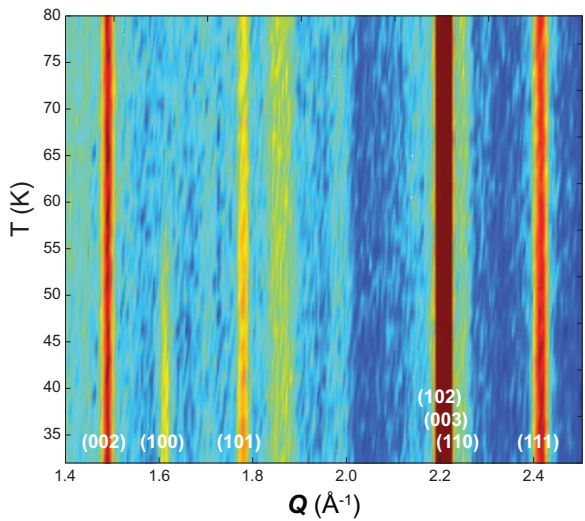}
\caption{(Color online) Variable temperature neutron diffraction surface across $T_\mathrm{N}$, showing the appearance of the
structurally forbidden (100) reflection and the increase in intensity of the (101) and (111) reflections due to magnetic ordering.}
\label{fig:mag_order}
\end{figure}

Information about the magnetic structure was obtained from neutron powder diffraction. On cooling below $\sim57$~K additional Bragg reflections appeared (Figure~\ref{fig:mag_order}). Their absence in the high resolution x-ray powder diffraction data, and their temperature dependence reflects their magnetic origin.

Rietveld modeling of the neutron data were performed using the GSAS program\cite{Larson:2004} with EXPGUI\cite{Toby:2001:J.Appl.Crystallogr.}. The magnetic form factor assuming spherical symmetry ($j_0$ and $j_2$) and the bound coherent scattering length of $^{237}$Np$^{3+}$ (10.55 fm) were taken from the international tables\cite{I.S.Anderson:2006:}.  A Land\'{e} splitting factor ($g$) of 0.6 was used.  Several parts of the diffraction patterns were excluded, due to the aluminium container and other encapsulation and sample environment artifacts.

\begin{figure}[t!]
\centering
\includegraphics[width=0.45\textwidth]{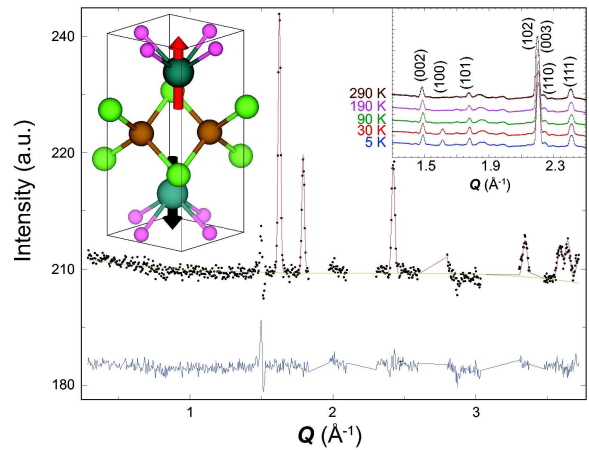}
\caption{(Color online) Plot of the Rietveld fit to the difference
of the $90$~K and $30$~K neutron powder diffraction data sets using a $P4/n'm'm'$ magnetic
model.  Portions of the difference spectra where strong nuclear
reflections led, due to thermal expansion, to strong up-down features
were excluded from the fit. The right inset shows the constant
presence of background features at all temperatures in the mid
momentum transfer region. The left inset shows the proposed
antiferromagnetic structure.} \label{NPD-refinement}
\end{figure}

The 5 K and 30 K data sets were fitted with the AFM $P4/n'm'm'$ model, whilst the remaining sets were fitted with a paramagnetic $P4/nmm$ model, the results of which are shown in Table I.  To confirm that all the magnetic Bragg intensities were being modeled correctly, a purely magnetic fit was carried out to the difference between the 30 K and 90 K data sets (Figure \ref{NPD-refinement}).  A list of reflections, with the measured and calculated structure factors are shown in Table \ref{tab:NPD_intensities}. These peaks indexed using the $P4/n'm'm'$ magnetic Space Group, correspond to the Np moments being aligned along the $c$-axis, with a ferromagnetic coupling within the basal plane, and an antiferromagnetic coupling between the planes (see the inset to Figure~\ref{NPD-refinement}), reminiscent of the low temperature ordering of the Sm moments in SmFeAsO\cite{Nandi,Ryan}.

No Fe moment is observed within the limits of the experiment. At 5 K, the ordered magnetic moment of the Np ions is of $1.70\pm0.07\mu_\mathrm{B}$, in excellent agreement with M\"{o}ssbauer spectroscopy estimations \cite{Gaczynski}.

\begin{table*}[h!t!]
  \caption{Crystallographic parameters as a function of temperature derived from Rietveld refinement of the structural model to match the neutron powder diffraction data from D20. ZrCuSiAs structure, space group $P4/nmm$ (origin choice 2) with Np, Fe, As, and O on the 2\textit{c}, 2\textit{b}, 2\textit{c}, and 2\textit{a} Wyckoff sites, respectively. The Np magnetic moment (Np $m_z$) refined in the $P4/n'm'm'$ magnetic space group in the low temperature phase is also presented here.}
  \label{tab:NPD_D20}
  \begin{tabular}{llllll}
    \hline
    $T$ / K                         & 5             & 30            & 90            & 190           & 290           \\
    \hline
    $a$ / \AA                       & 3.85525(15)   & 3.85512(15)   & 3.85530(15)   & 3.85851(16)   & 3.86235(17)   \\
    $c$ / \AA                       & 8.3374(6)     & 8.3371(6)     & 8.3324(6)     & 8.3466(7)     & 8.3663(7)\\
    Np $z$                          & 0.1507(6)     & 0.1508(6)     & 0.1507(6)     & 0.01512(7)    & 0.1519(8)\\
    As $z$                          & 0.6742(11)    & 0.6742(10)    & 0.6755(10)    & 0.6749(10)    & 0.6738(12)\\
    Np $U_\mathrm{iso}$ / \AA$^2$   & 0.028(2)      & 0.029(2)      & 0.027(2)      & 0.031(3)      & 0.036(3)\\
    Fe $U_\mathrm{iso}$ / \AA$^2$   & 0.032(3)      & 0.032(3)      & 0.033(3)      & 0.034(3)      & 0.034(3)\\
    As $U_\mathrm{iso}$ / \AA$^2$   & 0.032(4)      & 0.032(4)      & 0.035(4)      & 0.037(4)      & 0.038(4)\\
    O $U_\mathrm{iso}$ / \AA$^2$    & 0.028(5)      & 0.029(5)      & 0.027(4)      & 0.033(5)      & 0.039(5)\\
    Np $m_z$ / $\mu_\mathrm{B}$     & 1.70(7)       & 1.66(7)       & ---           & ---           & ---    \\
    $R_\mathrm{wp}$                 & 0.0212        & 0.0209        & 0.0205        & 0.0201        & 0.0198\\
    $R_\mathrm{p}$                  & 0.0155        & 0.0152        & 0.0151        & 0.0146        & 0.0141\\
    \hline
  \end{tabular}
\end{table*}

It is worth noting, that for the parent \textit{R}FeAsO ``1111'' compounds, the magnetic rare-earth moments order with the moments varying from 0.83 to 1.55 for CeFeAsO and NdFeAsO, respectively \cite{magnetism-Fe-sc-review}. The highest N\'{e}el temperature, of the rare-earth sublattice, in the \textit{R}FeAsO ``1111'' series, $T_\mathrm{N}$(Pr)=14 K observed for PrFeAsO \cite{Zhao08}, is four times lower than $T_\mathrm{N}=57 K$ estimated for NpFeAsO.

%The refined ordered magnetic moment on the Np ions is $1.891\pm0.013\mu_\mathrm{B}$. Due to the necessary encapsulation of our radioactive sample, regrettably the quality of the data does not allow us to infer any information regarding any ordering of the iron moments, with the exception of placing a upper limit on any ordered moment of $0.4\mu_\mathrm{B}$.

To explore further the electronic and magnetic character of NpFeAsO
we have performed first-principle local spin density approximation
(LSDA) and LSDA plus Coulomb-U (LSDA+U) calculations.
%Commonly
%accepted values for the Np atom Coulomb $U = 3$~eV and exchange $J =
%0.61$~eV were used.
In all calculations we used the crystal structure parameters determined experimentally at room temperature, assuming the
tetragonal ZrCuSiAs structure. Gaining inspiration from the neutron
scattering data, we assumed that the magnetic and crystallographic
unit cells coincide. Furthermore, we assumed that the magnetic
moment is aligned along $c$-[001]-axis, and considered non-magnetic
(NM), ferromagnetic (FM), and anti-ferromagnetic (AF) arrangements
for the Np moments. There were no magnetic moments set initially on the Fe
atoms. In all calculations we assumed stoichiometric NpFeAsO and did not
consider any defects or impurity NpO$_2$ phase. The effect of defects and impurities is left for further consideration.
% and checker-board AF moments arrangement for Fe atoms.

We used an in-house implementation of the full-potential linearized
augmented plane wave (FP-LAPW) method \cite{shick99}. This FP-LAPW
version includes all relativistic effects (scalar-relativistic and
spin-orbit coupling), and relativistic implementation of the
rotationally invariant LSDA+U \cite{shick01}. In the FP-LAPW
calculations we set the radii of the atomic spheres to
2.75~a.u.~(Np), 2.2~a.u.~(Fe, As), and 1.6~a.u.~(O). The parameter
$R_{Np} \times K_{\text{max}}=9.625$ determined the basis set size,
and the Brillouin zone (BZ) sampling was performed with 405
$k$~points.  For the neptunium $f$~shell, Slater integrals of $F_0 =
3.00$~eV, $F_2=7.43$~eV, $F_4=4.83$~eV and $F_6= 3.53$~eV were
selected to specify the Coulomb interaction~\cite{KMoore2009}. They
correspond to commonly accepted values for Coulomb $U = 3$~eV and
exchange $J = 0.61$~eV parameters. The dependence of Np-atom $m_J$
on the choice of Coulomb-$U$ in a range from 1 eV to 3 eV was
checked, and only small changes in the $m_J$ moment value were
found.

In LSDA, we found that spin-polarization decreases the total energy
with respect to the non-magnetic solution by 0.351 eV/ per formula
unit (f.u.) for the FM-solution, and by 0.335 eV/f.u. for the
AF-solution, suggesting an FM-ordered ground state. However, including the Coulomb-$U$ and exchange-$J$ in the fully-localized-limit (FLL) LSDA+U method, we obtain a total energy for the FM solution which is higher by 12 meV/f.u. than in the AF solution; on the contrary, if the double
counting correction is made in the around-mean-field-limit (AMF), the AF solution
becomes lower in energy than the FM one by 61.2 meV/f.u. Thus, both
flavors of LSDA+U yield an AF-ordered ground state, in agreement
with the experiment.

The  spin $m_S$, orbital $m_L$, and total $m_J$ magnetic moments for
the AF calculations with LSDA and LSDA+U are shown in
Table~\ref{tab:2}. The staggered local magnetic moments, which are due to magnetic polarization of the
Np $f$-shell are formed
at the two Np atoms.
%(see Fig. 4(a) where charge and spin density
%distributions near Np atoms in the unit cell are shown).
Comparison with the experimentally determined moment of $1.7$~$\mu_\mathrm{B}$
shows that LSDA is failing completely to reproduce the results of
neutron scattering experiments, whilst the AMF-LSDA+U calculations
overestimate the $m_J$ moment  value. On the other hand, the
FLL-LSDA+U calculations yield an $m_J$ value in reasonable
agreement with the experimental data, and the ratio of the orbital to total moment gives a \textit{g}-value of 0.58,
consistent with that for a free Np$^{3+}$ ion, and the value of 0.6 used in the NPD refinements.

It should be noted that when the total energy difference between the
FM and the AF solutions is used to make a "naive" molecular-field
theory estimate of the N\'eel temperature $T_\mathrm{N}$, the FLL-LSDA+U yields
$T_\mathrm{N} \sim 47$~K in reasonable agreement with the experimental data.
Thus we assume that the LSDA+U model with the FLL-double-counting
choice gives the most appropriate description of the NpFeAsO.

\begin{table}[b]
  \caption{Observed, $\mathrm{F_{o}^2}$, and calculated, $\mathrm{F_{c}^2}$, structure factors for the magnetic Bragg reflections in the $30 - 90$ K refinement.  Structure factors have been corrected for scale and extinction.}
  \label{tab:NPD_intensities}
  \begin{tabular}{lcll}
    \hline
    $hkl$ & $\bm{Q}$ (\AA$^{-1}$)     & $\mathrm{F_{o}^2}$ & $\mathrm{F_{c}^2}$\\
    \hline
100  &   1.63   &  0.9312            &        0.9128\\
101  &   1.80   &  0.2374            &        0.2480\\
102\footnotemark[1]  &   2.22   &  ---                 &        0.04335\\
111  &   2.43   &  0.4749            &        0.4442\\
112\footnotemark[1]  &   2.75   &  ---            &        0.4174\\
103\footnotemark[2]  &   2.79   &  0.09241            &        0.2043\\
113\footnotemark[1]  &   3.23   &  ---              &        0.02390\\
201  &   3.35   &  0.3697            &        0.3177\\
104\footnotemark[1]  &   3.43   &  ---              &        0.06999\\
202  &   3.59   &  0.3496            &        0.3294\\
210  &   3.64   &  0.3925           &       0.4317\\
211\footnotemark[2]  &   3.72   &  0.1687           &       0.1349\\
    \hline
  \end{tabular}
   \footnotetext[1]{Excluded region.}
   \footnotetext[2]{Partially excluded region.}
\end{table}

Although no magnetic polarization on other atoms in the unit cell was
initially assumed, the presence of the exchange splitting at the Np
atoms leads to induced staggered spin moments on the Fe atoms in
a checker-board AF arrangement. These are fairly negligible in LSDA,
but reach $\sim$ 0.1-0.25 $\mu_\mathrm{B}$ values in LSDA+U. Such a
moment on the iron was not visible in our neutron powder diffraction
measurements, but can not be excluded due to the background arising from the sample
encapsulation and to the high absorption cross-section of Np.
%The corresponding charge and spin density
%distributions are shown in Fig. 4(a).

\begin{table*}[htbp]
  \begin{tabular}{ccccccccccccccccc}
         \hline
 &\multicolumn{4}{c}{\bf  LSDA} &&\multicolumn{4}{c}{\bf AMF-LSDA+U} &&
 \multicolumn{4}{c}{\bf  FLL-LSDA+U}&&{\bf NPD} \\
            & Np & Fe & As & O & & Np & Fe & As & O & & Np & Fe & As & O & & Np \\
           \hline
 $m_S$&2.91 & 0.00 & 0.03 & 0.00 & &1.53 &0.25 & 0.01 & 0.00& &2.98&0.10&0.02&0.00& &2.26 \\
 $m_L$&-3.17& 0.00 & 0.00  & 0.00 & &-4.23&0.02 & 0.00 & 0.00& &-5.03&0.01&0.00&0.00& &-3.96 \\
 $m_J$&-0.25& 0.00  & 0.03  &0.00&&-2.70&0.27 & 0.01 & 0.00& &-2.05&0.11&0.02&0.00& &-1.7 \\
           \hline
  \end{tabular}
    \caption{Spin $m_S$, orbital $m_L$, and total $m_J$ magnetic moments ($\mu_\mathrm{B}$),
      for one (of two) Np, Fe, As, and O atoms in anti-ferromagnetic NpFeAsO
    resulting from calculations with LSDA and LSDA+U=3 eV with AMF- and FLL-double-counting flavors.
    The last column (NPD) provides the values obtained from the neutron powder diffraction studies
    making use of the \textit{g}-value of for a free Np$^{3+}$ ion.}\label{tab:2}
\end{table*}

\begin{figure}[htbp]
\centering
    \includegraphics[angle=270,width=0.475\textwidth,clip]{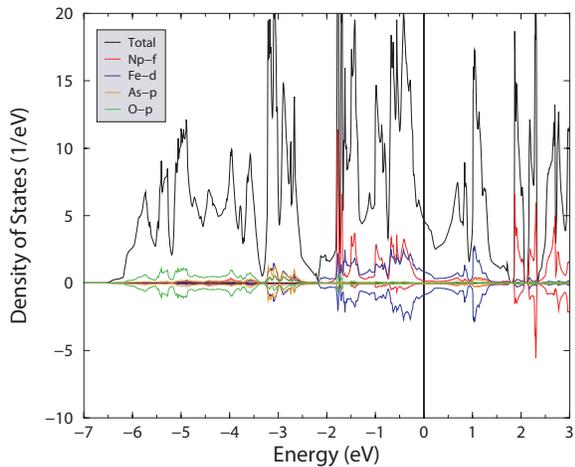}\caption
{Electronic structure calculations for NpFeAsO: the relativistic spin- and
orbital-resolved DOS (per unit cell) including the partial DOS}\label{fig:DOS}
\end{figure}

The total and partial (atom, spin and orbital-resolved) density of
states (DOS)  are shown in Fig.~5 calculated with FLL-LSDA+U= 3 eV.
The DOS near the Fermi energy ($E_F$) has mostly Fe-$d$ character,
while the As-$p$ and O-$p$ states are mostly located at 2-7 eV
energy interval below $E_F$. The Np-$f$ states are split by the exchange
interaction. The resulting DOS at $E_F$ of 4.8 states/eV corresponds
to the non-interacting value of the Sommerfeld coefficient $\gamma$= 5.6
mJ mol$^{-1}$K$^{-2}$ which is substantially lower than the experimental
value of 75 mJ mol$^{-1}$K$^{-2}$. However, the theoretical value of the
Sommerfeld coefficient is expected to increase due to electron mass
enhancement caused by dynamical electron interactions and
electron-phonon coupling.

The band structure and the Fermi surface (FS)  are shown in Fig.~6.
The FS consists of five sheets, each of them doubly degenerate.
Examination of the band structure shows that FS-1-3 sheets are
hole-like, and centered at the $\Gamma$-point. The FS-4 and FS-5 are
electron-like and centered at the M-A line. Note the fairly two
dimensional character of  the FS, and the strong resemblance to those previously presented for \textit{R}FeAsO\cite{Carrington,Singh}. It is seen that most of the states in the vicinity of the FS are located near $\Gamma$-$[0,0,0]$ and
$M$-$[\pi,\pi,0]$ k-points in the BZ suggesting the possibility of
$s\pm$ superconducting pairing mechanism\cite{Chubukov}.

A striking feature of NpFeAsO which makes it different from the
rare-earth-based counterparts, is the absence of the orthorhombic
structural distortion associated with the magnetic ordering.  In
order to understand this, we performed magnetic anisotropy energy
(MAE) calculations, rotating  the staggered AF magnetization from
the $c$-axis to the $a$-axis direction. What we found is that the
total energy $E_{tot}$ difference $E_{tot}(a)-E_{tot}(c)$ between
these two directions of the  magnetization is of  $\approx$  30.0
meV/f.u. This means that there is a strong positive uniaxial  MAE in
NpFeAsO. This MAE keeps the staggered AF magnetization along the
tetragonal $c$-axis and assists in the prevention of any
distortion in the ${a-b}$-plane.

Finally we turn to the discussion of the negative thermal expansion
(NTE). The most recent theory of the Invar effect~\cite{khmelevskyi}
is based on the use of a disordered local moment (DLM)
approximation. The paramagnetic state in the DLM approximation is
treated as a disordered pseudo-alloy with equal concentration of
randomly oriented "up" and "down" local moments. It is usually
implemented by making use of alloys theory in a coherent-potential
approximation. Regrettably, it is not currently possible to use the DLM
approximation together with the FP-LAPW basis used in this work.

\begin{table}[htbp]
  \label{tab:3}
  \begin{center}
  \begin{tabular}{cccccccccccc}
         \hline
 \multicolumn{3}{c}{Method}&\multicolumn{3}{c}{\bf  LDA} &\multicolumn{3}{c}{\bf  LDA+HIA} & \multicolumn{3}{c}{\bf  FLL-LSDA+U} \\
 \multicolumn{3}{c}{$V_{eq}$, ({$\AA$}$^3$)}&\multicolumn{3}{c}{111.50}& \multicolumn{3}{c}{116.11}&\multicolumn{3}{c}{119.92} \\
 \multicolumn{3}{c}{$B$, (MBar)}&\multicolumn{3}{c}{1.92}&\multicolumn{3}{c}{1.86}&\multicolumn{3}{c}{1.79} \\
           \hline
  \end{tabular}
  \end{center}
    \caption{The equilibrium volume $V_{eq}$ ({$\AA$}$^3$) for NpFeAsO resulting from paramagnetic LDA+HIA and
    anti-ferromagnetic LSDA+U calculations.}
\end{table}

\begin{figure*}[t]
%\centering
\centerline{\includegraphics[width=0.35\textwidth,bb=0 0 595
    725,angle=-90,clip]{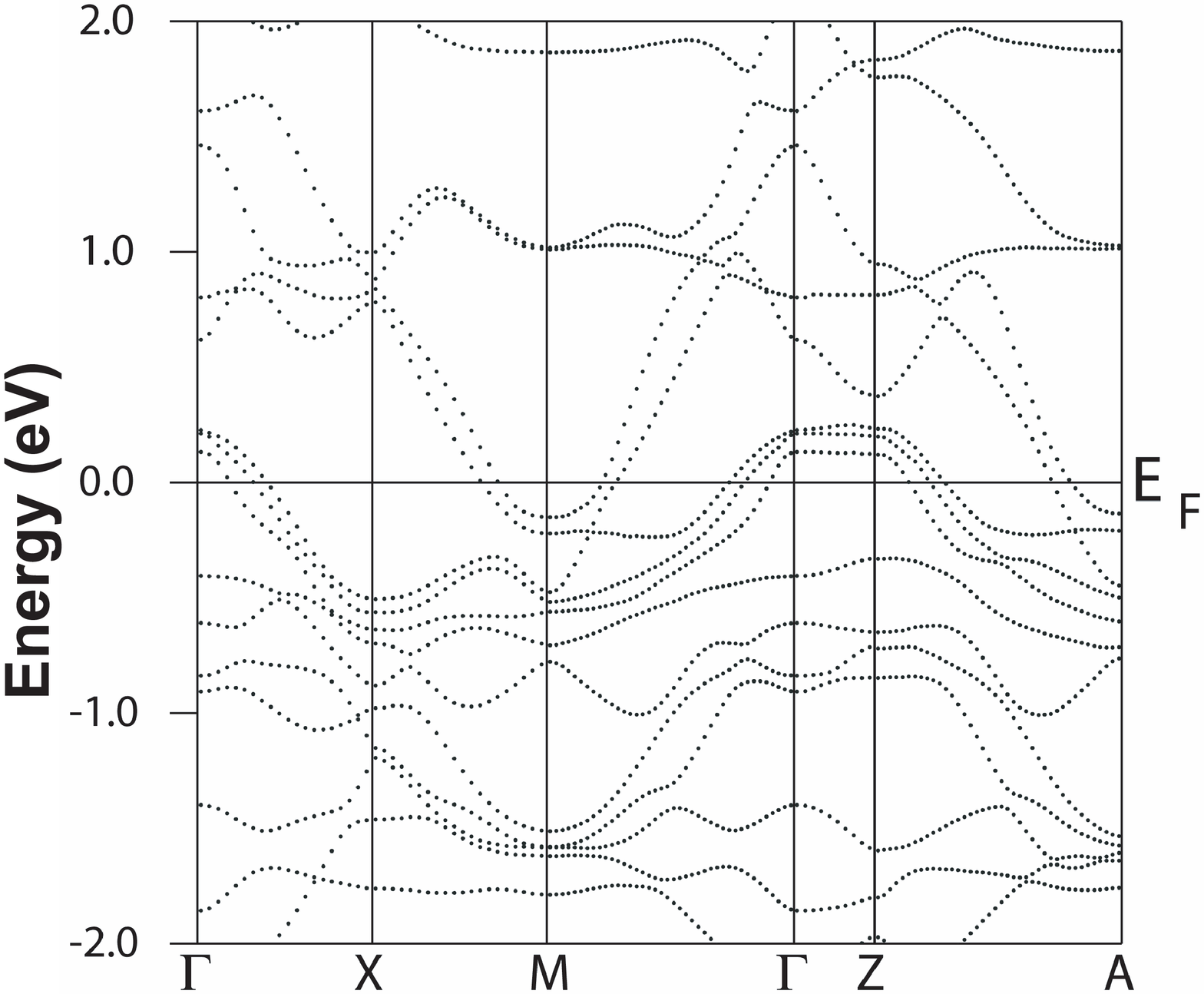}
\hspace{0.25cm}
\includegraphics[width=0.35\textwidth,bb=30 50 535
800,angle=-90,clip]{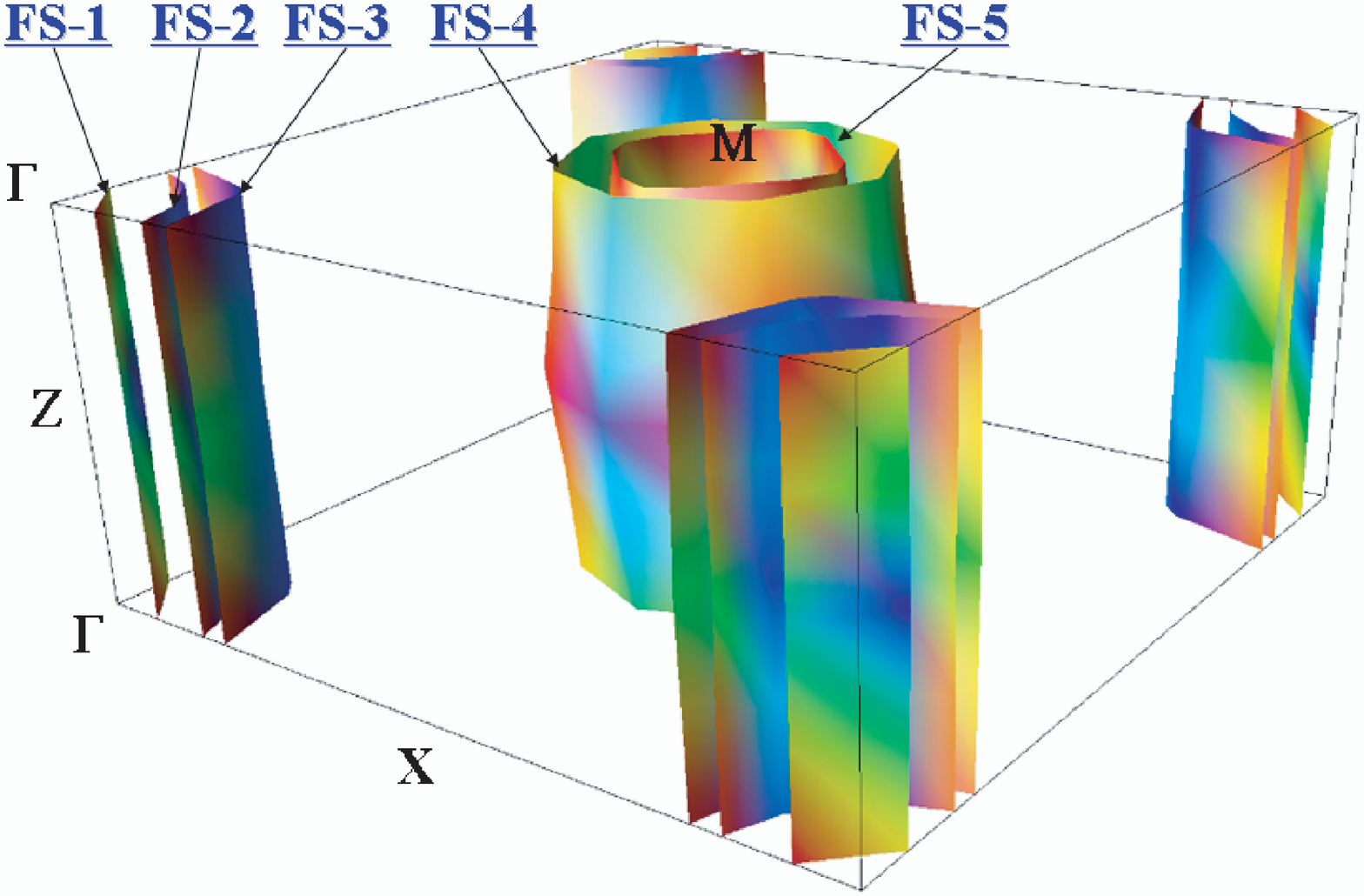}} \caption {(Left) Band
structure for NpFeAsO;(right) Fermi surface for NpFeAsO, calculated
making use of the FLL+LSDA+U=3 eV calculations.} \label{fig:Fermi}
\end{figure*}

Therefore, here the LDA+Hubbard I (HIA) approximation, as implemented in
Ref.\onlinecite{shick09}, is used to treat the Np-atom local moment
paramagnetic phase. The same values for the Slater integrals for Np
atom as in LSDA+U were used in LDA+HIA calculations (which correspond to %Commonly
a Coulomb $U = 3$~eV and an exchange $J = 0.61$~eV), and the inverse
temperature $\beta$=40 eV$^{-1}$ was chosen. The use of atomic-like
LDA+HIA is justified for NpFeAsO at high temperature since the
experiments presented above show a well localized character of the Np-atom
f-shell above the N\'{e}el temperature. Furthermore, the effective
paramagnetic moment of 2.84 $\mu_B$ per Np atom calculated in
LDA+HIA, is in very good agreement with the moment of
2.78$\pm$0.06~$\mu_B$ obtained from susceptibility measurements.

The total energy versus volume was calculated for LSDA+U and
LDA+HIA, and the equilibrium volumes ($V_{eq}$) were obtained using
a 3rd order polynomial fit. The $c/a$-ratio and the internal positions of Np and As atoms
were kept fixed in these calculations. The
resulting $V_{eq}$ for LSDA+U is reasonably close (with 3.3 \%
difference) to the experimental $V_{eq}= 124$ \AA$^3$ at T= 0 K,
confirming an adequate choice of the computational model for
NpFeAsO. However, the calculated $\omega$ of $3.1\%$, whilst
agreeing in sign, is about ten times bigger than the experimental value
($\omega=0.2\%$).

This quantitative disagreement between theory and experiment is
traced to the incorrect description of spin fluctuations on the Fe
atoms in LDA+HIA calculations. The Hubbard-I correction is included
for the Np atom only, and describes the local moment paramagnetic
phase in Np-sublattices. The Fe-atom sublattice is treated within
LDA, with no spin polarization (zero magnetic moment). This
approximation is not sufficient for the Fe-atom sublattice in the
paramagnetic phase of NpFeAsO. Note that it bears close similarities
to what has been found in Ref.~\onlinecite{turek}, for the $\omega$
calculations in ferromagnetic RE(=Gd,Dy,Er)Co$_2$ compounds. The
FM-state was treated with the so-called "open-core" approximation for
RE-atom, and the DLM approximation was used for PM-phase. The
calculated $\omega$ was found to be about five times bigger than in the
experiment, and no spin polarization on the Co-atoms was found in the
DLM calculations. By forcing a FM-moment on the Co-atom
sublattice (applying DLM to the RE-sublattice only), the results
improved quantitatively, and the value of $\omega$ reduced
substantially.

\section{Conclusions}

In conclusion, we successfully synthesized NpFeAsO, one of the few actinide-based iron-oxypnictide compound (the other reported
oxypnictides containing actinide elements are U$_2$Cu$_2$As$_3$O\cite{Kaczorowski} , UCuPO \cite{UCuPO,Wells}, and
NpCuPO \cite{Wells}). Although one may expect the physical properties of NpFeAsO to be similar to the lanthanide 1111
analogues, in reality this system behaves differently. In particular, we do not observe any structural transition, which is in agreement with the fact that neither transport properties (resisitivity and Hall effect) nor specific heat measurements reveal high temperature anomalies associated with an in-plane magnetic ordering in the Fe-As layer. We suggest that the lack of the orthorhombic structural distortion is caused by a strong positive uniaxial magnetic anisotropy energy.

Our results reveal that NpFeAsO exhibits an anomalous thermal expansion which is intimately linked to a giant magneto-expansion
occurring as the system undergoes antiferromagnetic ordering. Negative thermal expansion was observed in PrFeAsO\cite{Kimber1}, and we believe that it is a common feature in the 1111 family with a magnetic trivalent ion. However the effect in NpFeAsO is at least 20 times stronger and is clearly associated with the antiferromagnetism of the Np sublattice.

This invar effect is usually observed in transition metal disordered alloys (e.g. Fe-Ni,Fe-Pt~\cite{Wassermann}), and rare-earth ferromagnets (e.g. GdCu$_2$~\cite{lindbaum}) and anti-ferromagnets (e.g. PrFeAsO~\cite{Kimber1}). It was also found in $\alpha$-U metal \cite{Barret} and later this effect was explained by the onset of a charge density wave. Even more striking is the observation of a NTE in $\delta$- Pu\cite{Jette}, where the absence of magnetic ordering necessitates an alternative mechanism to that based on two spin states with different volumes, as for invar alloys \cite{Guillaume,Weiss}. Instead in that case it is proposed that the negative thermal expansion is a signature of criticality arising from proximity to a localization-delocalization transition\cite{Savrasov}. Although uranium and plutonium have rather overshadowed their neighbor, neptunium\cite{Ibers}, here, in reporting the observation of NTE in NpFeAsO, neptunium will finally take centre-stage. To the best of our knowledge, NpFeAsO is the first actinide-based antiferromagnetic material, in which negative thermal expansion has been observed. This behavior originates from the magnetic order, but more exotic explanations, e.g. possible criticality at the interface between localised and itinerant electrons, should also be taken into consideration.

%It should be added that that negative thermal expansion and anisotropic magneto-elastic coupling is often observed in rare-earth based antiferromagnets \cite{rotter05}.

%In conclusion, our results reveal that NpFeAsO exhibits an anomalous thermal expansion which is intimately linked to a giant magnetoexpansion occurring as the system undergoes antiferromagnetic ordering.

%\section{Supplementary Methods}

%\noindent \textbf{} \vspace{5mm}

\section*{Acknowledgement}
This work has been performed at the Institute of Transuranium Elements within its "Actinide User Laboratory" program, with financial support to users provided by the European Commission. Np metal required for the fabrication of the compound was made available through a loan agreement between Lawrence Livermore National Laboratory and ITU, in the frame of a collaboration involving Lawrence Livermore National Laboratory, Los Alamos National Laboratory, and the US Department of Energy. 

The support from Czech Republic Grants GACR P204/10/0330, and GAAV IAA100100912 is thankfully acknowledged. Work at Princeton supported by US DOE grant DE FG02-98ER-45706. TK acknowledges the European Commission for financial support in the frame of the "Training and Mobility of Researchers" programme.

Many thanks to everyone who helped with the experiments and preparation of this paper, in particular F. Kinnart (ITU), A. Hesselschwerdt (ITU) and  L. Havela (Charles University, Prague).

\bibliographystyle{apsrev}
\bibliography{nfao_nm}% Produces the bibliography via BibTeX.

\end{document}